\documentclass[preprintnumbers,superscriptaddress,amsmath,floatfix,amssymb,prd]{revtex4}
\usepackage{amsmath,amssymb,bm}
\usepackage{graphicx}
\usepackage{hyperref}
\usepackage{amsfonts}
\usepackage{latexsym}
\usepackage{amsmath}
\usepackage{amssymb}
\usepackage{slashed}
\usepackage{longtable}
\usepackage{yfonts}
\usepackage{color}
\newcommand{\beq}{\begin{eqnarray}}
	\newcommand{\eeq}{\end{eqnarray}}

\begin{document}

\title{Hydrodynamic Waves in an Anomalous Charged Fluid}

\author{Navid Abbasi}

\affiliation{School of Particles and Accelerators, Institute for Research in Fundamental Sciences (IPM), P.O. Box 19395-5531, Tehran, Iran}

\author{ Ali Davody}
\affiliation{School of Particles and Accelerators, Institute for Research in Fundamental Sciences (IPM), P.O. Box 19395-5531, Tehran, Iran}

\author{ Kasra Hejazi}
\affiliation{Department of Physics, Sharif University of Technology, Tehran, Iran}

\author{ Zahra  Rezaei}
\affiliation{Department of Physics, Tafresh University, P. O. Box 79611-39518, Tafresh, Iran
}
\affiliation{School of Particles and Accelerators, Institute for Research in Fundamental Sciences (IPM), P.O. Box 19395-5531, Tehran, Iran}


\begin{abstract}
	{We study the collective excitations in a relativistic fluid with an anomalous  $U(1)$ current. In $3+1$ dimensions at zero chemical potential, in addition to   ordinary sound modes we find two propagating modes in presence of an external magnetic field.
	The first one which is a transverse degenerate mode, propagates with a velocity proportional to the coefficient of gravitational anomaly; this is in fact  the Chiral Alfv\'en wave recently found in \cite{Yamamoto:2015ria}.   Another one  is a wave of density perturbation, namely a chiral magnetic wave (CMW). The velocity dependence of CMW on the chiral anomaly coefficient is  well known. We compute the dependence of CMW's velocity on the coefficient of gravitational anomaly as well. We also show that the dissipation splits the degeneracy of CAW.
    At finite chiral charge density we show that in general there may exist five chiral hydrodynamic waves. Of these five waves, one is the CMW while the other four are mixed Modified Sound-Alfv\'en waves. It turns out that in propagation transverse to the magnetic field no anomaly effect appear while in  parallel to the magnetic field we find sound waves become dispersive due to anomaly. }
\end{abstract}
\maketitle


\section{Introduction}

\label{1}

Study of fluids with broken parity symmetry has attracted much attention recently. Parity may be broken in the system due to the presence of either an external magnetic field or a rotation in the fluid.  Currents along the direction of an external magnetic field discussed earlier in \cite{Vilenkin:1980fu}, has been recently argued to be realized in heavy-ion collisions \cite{Kharzeev:2007jp,Fukushima:2008xe}. This is termed as the chiral magnetic effect, CME. Analogously, chiral vortical effect, CVE, is related to currents in the direction of a rotation axis \cite{Vilenkin:1979ui}. The vorticity term which is responsible for this effect in the fluid constitutive current has	been  discovered in the context of gauge-gravity duality \cite{Erdmenger:2008rm,Banerjee:2008th}. 

The long time missed vorticity term seems to be in contradiction with
 existence of a positive divergence entropy current. However, because the parity violating terms like vorticity violate time-reversal as well, one may expect their associated transport coefficients to be non-dissipative. Considering the latter fact, Son and Soruka showed that the vorticity term is not only allowed by symmetries, but also is required by the triangle anomalies and the second law of thermodynamics \cite{Son:2009tf}. They computed the coefficients of both CME and CVE terms in terms of the anomaly coefficient at non-zero chemical potential ($\mu \ne 0$). These so-called anomalous transport coefficients vanish at zero chemical potential. The non-vanishing contribution to anomalous transport at $\mu=0$ was firstly observed in \cite{Bhattacharya:2011tra} and then was computed in \cite{Neiman:2010zi,Landsteiner:2011cp} by considering the mixed gravitational anomaly in $3+1$ dimensions.  
 
 The issue has  been also developed through other approaches\footnote{ The Chiral effects have been also studied in Lifshitz hydrodynamics  \cite{Roychowdhury:2015jha}.}. 
 For example, a  new kinetic theory containing such effects has been derived from the underlying quantum field theory \cite{Son:2012wh,Stephanov:2012ki}. It has been shown that the Berry monopole is responsible for the CME and CVE  \cite{Stephanov:2012ki,Chen:2012ca}.
Chiral magnetic effect has been also  studied in the context of lattice field theory \cite{Buividovich:2009wi,Buividovich:2010tn}.

Non-dissipative character of anomalous transport has been discussed by some authors. Apart from explanations based on symmetry \cite{Son:2009tf,Kharzeev:2011ds},
it has been recently illustrated with an example in the context of gauge-gravity duality. Computing the drag force exerted on a heavy quark moving in a general parity violating fluid \footnote{In the context of gauge-gravity duality, the gradient corrections to the drag force has been computed for the first time in \cite{Abbasi:2012qz,Abbasi:2013mwa}.}, the authors of \cite{Rajagopal:2015roa} have found a particular setting in which the CME- or CWE-induced current flows past the heavy quark without exerting any drag force on it. 

On the other hand, a usual way to study the transport phenomena is to investigate the long wave-length fluctuations around equilibrium state of the fluid.  Associatively, non-dissipative nature of the anomalous transport coefficients may be better understood via studying the hydrodynamic excitations in the chiral fluid. So in  this paper we consider a  fluid of chiral particles, i.e. single right-handed fermions, and compute the spectrum of its collective excitations to first order in derivative expansion. 

 Let us recall that in a parity preserving fluid in $3+1$ dimensions the only collective modes  are the two ordinary sound modes. However when taking into account the effect of dissipation, one finds four hydrodynamic modes in a charged fluid at zero chemical potential \cite{Kovtun:2012rj}.  Of these modes, two are the dissipating sound modes while the other two are pure shear modes. In \cite{Abbasi:2015nka} we showed that in presence of an external magnetic field, one of the latter shear modes would split into two new shear modes. As a result, one finds that the dissipation, when accompanying with presence of  magnetic field, excites all five possible hydrodynamic modes corresponding to five microscopic conserved charges in the system. 
 
In the current paper we study the hydrodynamic excitations in a parity violating fluid in presence of a background magnetic field. Our study includes two parts. Firstly, we consider a system at zero chemical potential in $3+1$ dimensions and compute the hydrodynamic modes in the absence of dissipation. we find four distinguished  modes as it follows: two longitudinal sound modes and two  chiral modes. The appearance of chiral modes are due to presence of chiral anomaly as well as the gravitational anomaly. Of these two, the chiral wave with a velocity proportional to the gravitational anomaly coefficient is the
so-called Chiral Alfv\'en, recently found by Yamamoto too\cite{Yamamoto:2015ria}; this  mode is a  wave of momentum fluctuations. Another chiral wave that we obtain,   is nothing but a  CMW. The dependence of  CMW velocity on the chiral anomaly is well known. We find the dependence of CMW velocity on the  gravitional anomaly coefficient as well. To do so, we  use the anomalous transport coefficients including the effect of gravitational anomaly as well as the chiral anomaly effects, in  Landau frame.

One may expect in a dissipative chiral fluid one of the above-mentioned four modes to be split into two dissipative waves. We show that this actually happens for the  Alfv\'en wave.
In summary, in 3+1 dimensions, five distinguished hydrodynamic modes may be excited in a dissipative chiral fluid: two dissipating sound modes and three dissipating chiral waves.

Another part of our results is related to the hydrodynamic waves in a chiral fluid at finite density. In reality, such fluid might exist above the electroweak phase transition where the $SU(2) \times U_Y(1)$ symmetry is not broken. Since the hypermagnetic field associated with the $U_Y(1)$ couples differently to right- and left-handed electrons,  the high-
temperature plasma there is chiral \cite{Giovannini:1997eg}. We will explore hydrodynamic fluctuations in such plasma. We will show that in this regime, sound modes specifically, will be modified remarkably. They become mixture of longitudinal and transverse waves which one may refer to as the modified sound waves. Depending on the relative situation of magnetic field and wave vector, we carefully compute hydrodynamic modes in different cases.

 As it is well known, the chiral anomaly is present in even space-time dimensions. In $1+1$ dimensions, the anomalous transport has been discussed in the context of effective field theory. Using the second law of thermodynamics, the authors of \cite{Dubovsky:2011sk} derived a formula for the only anomalous transport coefficient in $1+1$ dimensions in terms of thermodynamic functions.  The same relation has been also obtained in 
\cite{Jain:2012rh} from the partition function. 
 
 To complete our discussion,  we also compute the hydrodynamic fluctuations of a non-dissipative chiral fluid in $1+1$ dimensions. In addition to  two ordinary sound waves, we find a new propagating wave; the so-called "one-and-a-halfth sound" which was previously found in \cite{Dubovsky:2011sk} from the effective field theory method. Compared to the earlier results, we explicitly compute the velocity of first mode in terms of anomaly
 coefficient\footnote{ Two dimensional chiral transport has been also studied at weak coupling in \cite{David:2010qc,Chowdhury:2015pba}.}.
 
 The paper is organized as it follows: In section \ref{sec2}  we give a brief review of the parity odd fluid dynamics in $3+1$ dimension. We continue the topic by studying a neutral chiral fluid in \ref{sec3}. We first compute the hydro modes and their amplitudes and then physically interpret them. In \ref{sec4}, we first study a charged fluid with no anomaly.  Then for different relative situations of wave vector and magnetic field, we compute the hydrodynamic modes of a chiral charged fluid.
 In \ref{sec5}, we study the effect of anomaly on the collective excitations of a 1+1 dimension chiral fluid. After mentioning some follow up questions in   \ref{App}, we give some comments about the collective excitations in a parity violating fluid in 2+1 dimensions.

\section{Parity violating fluid in $3+1$ dimensions	} \label{sec2}

Let us recall that the parity violating terms in $3+1$ dimensions have been shown to be associated with triangle anomaly of chiral currents.
In presence of a background gauge field $A_{\mu}$, 
the equations of hydrodynamic for a normal fluid with a conserved charge, with  $U(1)^3$ anomaly, take the form:
\begin{equation}
\begin{split}
\partial_{\mu}T^{\mu \nu}=&\,F^{\nu \lambda} J_{\lambda}\\
\partial_{\mu} J^{\mu}=&\, \mathcal{C} E_{\mu} B^{\mu}
\end{split}
\end{equation}
where we have defined the electric and magnetic field in the rest frame of the fluid  as $B^{\mu}=\frac{1}{2}\epsilon^{\mu\nu\alpha\beta}u_{\nu}F_{\alpha \beta},\,\,\,\,\,\,E^{\mu}=\,F^{\mu \nu}u_{\nu}$ \cite{Son:2009tf}.

The  energy-momentum tensor and the chiral current are 
 \begin{equation}\label{TJ}
 	\begin{split}
 		T^{\mu \nu}=& \,(\epsilon+p) u^{\mu} u^{\nu}+ p \,\eta^{\mu \nu} +\tau^{\mu \nu}\\
 		J^{\mu}=& \,n u^{\mu} +\nu^{\mu}.
 	\end{split}
 \end{equation}
Here the thermodynamic parameters $\epsilon(\mu,T)$, $p(\mu,T)$ and $n(\mu,T)$ are the values of energy density, pressure and charge density respectively in an equilibrium state.  
 The equilibrium state is specified with
 \begin{equation}
 u^{\mu}=(1,0,0,0),\,\,T=\text{Const.},\,\,\mu=\text{Const.},\,\,\boldsymbol{B}=\boldsymbol{0}
 \end{equation} 
 with the pressure $\bar{p}=\bar{p}(\mu, T)$ satisfying:
 \begin{eqnarray}
 d\bar{p}&= \bar{s} dT+ \bar{n} d\mu\\
 \bar{\epsilon}+\bar{p}&=\bar{s} T+\bar{n} \mu.\,\,\,\,
 \end{eqnarray}

 In the Landau-Lifshitz frame  where $u_{\mu}\tau^{\mu \nu}=0$ and $u_{\mu} \nu^{\mu}=0$ \cite{Landau,Bhattacharya:2011tra} we may write
 \begin{eqnarray}\label{disspart}
  \tau^{\mu \nu}=-\eta  P^{\mu\alpha}P^{\nu\beta}\left(\partial_{\alpha}u_{\beta}+\partial_{\beta}u_{\alpha} \right)-\left(\zeta-\frac{2}{3}\eta\right) P^{\mu \nu} \partial.u\\
  \nu^{\mu}=  - \sigma T P^{\mu \nu} \partial_{\nu}\left(\frac{\mu}{T}\right)+\sigma E^{\mu}+\xi \omega^{\mu}+\,\xi_B B^{\mu}\,\,\,\,\,\,\,\,\,\,\,\,\,\,\,\,\,\,\,\,\,\,
 \end{eqnarray}
 Here $\omega^{\mu}=\frac{1}{2}\epsilon^{\mu\nu\alpha\beta}u_{\nu}\partial_{\alpha}u_{\beta}$ is the vorticity and the $\xi$ and $\xi_{B}$ are the anomalous transport coefficients corresponding to CVE and CME \cite{Son:2009tf,Neiman:2010zi}
 \begin{eqnarray}\label{xi}
 \xi=\mathcal{C}\mu^2\left(1-\frac{2}{3} \frac{\bar{n} \mu}{\bar{\epsilon}+\bar{p}}\right)+\mathcal{D} T^2\left(1- \frac{2\bar{n} \mu}{\bar{\epsilon}+\bar{p}}\right)\\\label{xiB}
 \xi_B=\mathcal{C}\mu\left(1-\frac{1}{2} \frac{\bar{n} \mu}{\bar{\epsilon}+\bar{p}}\right)-\frac{\mathcal{D}}{2}  \frac{\bar{n} T^2}{\bar{\epsilon}+\bar{p}}\,\,\,\,\,\,\,\,\,\,\,\,\,\,\,\,\,\,\,\,\,\,\,\,\,\,\,\,
 \end{eqnarray}
 where $\mathcal{C}$ and $\mathcal{D}$ are the coefficients of chiral anomaly and gravitational anomaly respectively, as \cite{Gao:2012ix,Golkar:2012kb,Jensen:2012kj}
 \begin{equation}
 \mathcal{C}=\frac{1}{4 \pi^2},\,\,\,\,\,\,\,\mathcal{D}=\frac{1}{12}.
 \end{equation}
Let us note that firstly in \cite{Son:2009tf}, Son and Surowka computed the chirally anomaly contributions to $\xi$ and $\xi_{B}$, namely the statements in front of $\mathcal{C}$, and then the authors of \cite{Landsteiner:2011cp} generalized  \cite{Son:2009tf} by computing the contribution of gravitaional anomaly, namely $\mathcal{D}$ terms. 

Before ending this subsection let us briefly discuss how one can consider the background electromagnetic field consistent with the hydrodynamic expansion. 
Note that we take the strength of $A_{\mu}$ of the same order of the
temperature and the chemical potential, so $A_{\mu}\sim O(\partial^0)$ and  $F^{\mu \nu}\sim O(\partial)$. 

Let us recall that we are interested in the problem of how different wave-lengths behave in presence of an external magnetic magnetic field. It is natural to assume that the value of this magnetic field is constant. However this assumption becomes problematic when one considers the values of wave-lengths that are not of the same order as length-scale corresponding to the magnetic field ($\ell_{B}$). The reason is that when freely studying the wave-lengths much larger than $\ell_{B}$, additional not-necessarily-small contributions would arise which are not captured by the hydrodynamic expansion. It means that the assumption that $F^{\mu\nu}$ is of first order will be no longer valid when it is constant. 
Therefore, our study is restricted to the values of wave-lengths that are of same scale as $\ell_{B}$. 

Our strategy is to study the problem following the method used in   \cite{Buchbinder:2008dc}. We consider
a restricted interval of wave-lengths, which contains those of the same order as the $\ell_{B}$. We study only the values of wave-lengths inside this so-called window. 
 We then request the magnetic field to approach zero as the wave-vector tends to zero. To proceed one may consider the following relation between the magnetic field and the wave-vector \cite{Buchbinder:2008dc}:
$$\boldsymbol{B} = \tilde{\alpha} \,\boldsymbol{k}.$$
However, in contrast to  \cite{Buchbinder:2008dc}, $\tilde{\alpha}$ is a matrix here and thus the magnetic field and the wave-vector are not in general parallel.


%
 
In the next subsection, we compute the hydrodynamic modes around the equilibrium state  and show that how the simultaneous presence of dissipative effects and the anomalies may lead  to excite three dissipating chiral waves in the medium.

\section{Zero Chemical Potential}\label{sec3}
In order to study the hydrodynamic fluctuations, we should take the hydrodynamic equations and linearize them around the equilibrium state.
Instead of five usual hydrodynamic variables, namely one temperature field $T(x)$, one chemical potential field $\mu(x)$ and three components of velocity field $u^{\mu}(x)$, we would rather to take the following set of variables as the hydrodynamic variables:
$\phi_a=\left(T^{00}(x), T^{0i}(x),J^0(x)\right)$. The importance of this choice is that, to each of these hydrodynamic variables, a quantum operator corresponds. 
In the two following subsections, we compute the hydrodynamic modes around the equilibrium state firstly for the zero chemical potential case and then we physically interpret our results. 

\subsection{Hydrodynamic Modes}

 To first order in linear fluctuations, the super field $\phi_a$ may be written as 
$\label{variables}
	\phi_a=\,\big(\phi_0,\phi_i,\phi_5\big)=\,\big(\bar{\epsilon}+\delta \epsilon,\frac{}{}\pi_i,\, n\big)
$
where $\pi_i=\frac{v_i}{\bar{\epsilon}+\bar{p}}$ and $i=1,2,3$.
In terms of spacial Fourier transformed field $\phi_a(t, \boldsymbol{k})$, the linearized hydrodynamic equations are written as
 \begin{equation}\label{linearized3+1}
 \begin{split}
& \partial_{t} \delta \epsilon +\, i k^{j} \pi_{j}=0\\
& \partial_{t}\pi_{i}+\,i k_{i} v_s^2 \delta \epsilon   + \mathcal{M}_{i j } \pi_{j}=\,-i D F_{i m} k^{m}  n \,+\\
 &\,\,\,\,\,\,\,\,\,\,\,\,\,\,\,\,\,\,\,\,\,\,\,\,\,\,\,\,\,\,\,\,\,\,\,\,\,\,\,\,\,\,\,\,\,\,\,\,\,\,\,\,\,\,\,\,\,\,\,\,\,\,\,\,\,\,\,\,\,\,\,\,F^{im}\left(\frac{\sigma}{\bar{w}} F_{mj}+i\frac{ \xi}{2\bar{w}} \epsilon_{m l j}k^l\right)\pi^j\\
 & \partial_{t}  n+ \,\left(k^2 D-\frac{i}{2}\left(\frac{\partial \xi_B}{\partial n}\right)_{\epsilon}\epsilon^{ijm}F_{ij}k_m\right)  n +\frac{i  \sigma}{\bar{w}}  k_{j} F^{j k} \pi_{k}=0\\
  \end{split}
 \end{equation}
 Here $\mathcal{M}_{i j }=\gamma_{\eta}(\boldsymbol{k}^2 \delta_{i j}- k_{i} k_{j})+ \gamma_{s} k_{i}k_{j}$. Note that in the above equations, the anomalous transport coefficients have to be evaluated at zero chemical potential. While $\xi_B$ vanishes at $\mu=0$, its fluctuations will no longer vanish at the same limit
 \begin{eqnarray}\label{put}
 \xi &=&\,\mathcal{D}T^2\\
\left( \frac{\partial\xi_{B}}{\partial n} \right)_{\epsilon}&=&\,\frac{\mathcal
{C}}{\chi}-\frac{\mathcal{D}}{2}\frac{T^2}{\bar{w}}.
 \end{eqnarray}
 In what follows we first compute the hydrodynamic modes in terms of $\xi$ and $\frac{\partial \xi_B}{\partial n}$ and then, we re-express them in terms of the anomaly coefficients.
 
 In order to find hydrodynamic modes we use the super field notation $\phi_a$   
 and rewrite the linearized equations (\ref{linearized3+1}) as
\begin{equation}\label{eqsuperfield}
\partial_{t} \phi_{a}(t, \boldsymbol{k})+\,M_{a b}(\boldsymbol{k})\, \phi_{b}(t, \boldsymbol{k})=0
\end{equation}  
by introducing

\begin{widetext}
\begin{equation}
M_{a b}=
\left( {\begin{array}{ccc}
0 & i k_j & 0\\ 
i k^{i} v_{s}^2 & \mathcal{M}^{i}_{ j}-{\sigma}{\bar{w}}\big(B_j B^i- \boldsymbol{B}^2\delta_j^i\big)+\frac{ i \xi}{2\bar{w}}\big(\boldsymbol{B}.\boldsymbol{k}\delta_j^i-B_j k^i \big)  & -i D \epsilon^{inm}B_m k_n \\
	0    & -\frac{i \sigma}{\bar{w}}\epsilon_{jnm}B^n k^{m}  & \boldsymbol{k}^2 D+ i \left(\frac{\partial \xi_B}{\partial n}\right)_{T} \boldsymbol{B}.\boldsymbol{k}
	\end{array} } \right). 
\end{equation}
\end{widetext}
Hydrodynamic modes as the poles of response functions may be found via solving the following equation:
\begin{equation}
\det\left(-i \omega\delta_{ab}+\frac{}{} M_{ab}(\boldsymbol{k})\right)=0.
\end{equation}
Doing so, we find four hydrodynamic modes in the absence of dissipative effects
\begin{eqnarray}
\omega_{1,2}^{(0)}(\boldsymbol{k})&=&\,\pm v_{s} k\\
\label{first alfven}
\omega_{3,4}^{(0)}(\boldsymbol{k})&=&-\frac{\xi}{2\bar{w}} \boldsymbol{B}.\boldsymbol{k}\\
\label{first CMW}
\omega_{5}^{(0)}(\boldsymbol{k})&=& \left(\frac{\partial \xi_B}{\partial n}\right)_{\epsilon} \boldsymbol{B}.\boldsymbol{k}.
\end{eqnarray}
So far, we have only computed the dispersion relation of hydrodynamic modes to zero order in derivative expansion, namely $\omega^{(0)}$.
Just like a non-chiral fluid, we see that there exist two ordinary sound modes here. However, due to the effect of anomalies, two other hydrodynamic modes may propagate in  a chiral fluid. The first one, which is itself a degenerate mode, namely $\omega_{3,4}$, was recently found by Yamamoto too. This mode which is referred to as the Chiral Alfv\'en wave  in \cite{Yamamoto:2015ria}, propagates with a velocity that
goes to zero with the coefficient of gravitational anomaly (see table(\ref{tabelmodes3+1})). 
As was noted in \cite{Yamamoto:2015ria}, in an incompressible fluid, the Chiral Alfv\'en wave would be a transverse wave (We discus about this point in next subsection.).  The last mode, namely $\omega_{5}$, is a kind of well-known Chiral Magnetic Waves. In   \cite{Kharzeev:2010gd}, Kharzeev and Yee showed that the coupling
between the density waves of the electric and chiral charges leads to existence of a new type of gapless excitations in the plasma of chiral fermions. They called the latter as the Chiral Magnetic Wave.
As they emphasized in \cite{Kharzeev:2010gd}, the CMW could also exist in a fluid of single right handed fermions; the  mode $\omega_{5}$ given above is exactly the  CMW they pointed out (See next subsection.).
 \subsection{Mode Analysis}
 Using relations \ref{put}, we have re-expressed modes found in previous subsection in table(\ref{tabelmodes3+1}). In addition, we have written the normalized amplitudes of fluctuations in the last column of the table. Note that  $\delta \phi_{n}$ is the amplitude corresponding to the mode $\omega_{n}(k)$; in position space it may be written as
 \begin{equation}
\delta \phi_{n}(k,\omega_{n}) e^{i \boldsymbol{k.x}- i \omega_{n}(k)t}.
 \end{equation}  
 We know sound mode is the propagation of energy and momentum fluctuations. Since the vectorial propagating component of sound amplitudes $\delta \phi_{1,2}$  is in the direction of wave vector $\boldsymbol{k}$, sound would be a longitudinal mode expectedly. The situation is different with Chiral Alf\'en Waves. Both $\delta \phi_{3}$ and $\delta \phi_{4}$ are fully vector type fluctuations i.e. wave of momentum fluctuations. In contrast to sound modes, both of them are transverse \footnote{In the special case of incompressible fluid, Yamamoto showed that CAW is a transverse mode before\cite{Yamamoto:2015ria}}:
 \begin{equation}
 \begin{split}
 \boldsymbol{k.}\big(\hat{\boldsymbol{B}} \times \hat{\boldsymbol{k}}\big)&=0\\
 \boldsymbol{k.}\big(\hat{\boldsymbol{B}}-(\hat{\boldsymbol{B}}.\hat{\boldsymbol{k}}) \hat{\boldsymbol{k}}\big)&=0.
 \end{split}
 \end{equation}  
 
  The fifth mode in the table(\ref{tabelmodes3+1}) is the wave of scalar fluctuations, i.e. a density wave (CMW). This is why we refer to this mode as the Chiral Magnetic Wave.
  It is worth mentioning that the dependence of CMW's velocity on the chiral anomaly coefficient, is the same for both single right-handed (our case) and mixed chirality (\cite{Kharzeev:2010gd} case)  plasmas. Our result shows that the velocity of CMW may also depend on the gravitational anomaly coefficient, if one takes into account the effect of gravitational anomaly in computing the anomalous transport coefficients. 

\begin{longtable*}{|l|c|c|}
	\hline
Type of mode	& Dispersion relation & Amplitude   \\
	\hline
	\,\,\,\,\,\,\,\,sound  & $\omega_{1,2}^{(0)}(\boldsymbol{k})=\,\pm v_{s} k$ & $\delta \phi_{1,2}(k,\omega_{1,2}) = \left( \mp \frac{1}{\sqrt{\beta _1}},\hat{\boldsymbol{k}},0 \right) 
	$\\
   & &	\\
	\hline
	&&\\
Chiral 	Alfv\'en
	 &$\omega_{3,4}^{(0)}(\boldsymbol{k})=\,-\frac{\mathcal{D}}{2}\frac{T^2}{\bar{w}} \boldsymbol{B}.\boldsymbol{k}$  & $\delta \phi_{3}(k,\omega_3) = \left( 0,\hat{\boldsymbol{B}} \times \hat{\boldsymbol{k}},0 \right)$	  \\
	
	  &  & $\delta \phi_{4}(k,\omega_4) = \left( 0,\hat{\boldsymbol{B}}-(\hat{\boldsymbol{B}}.\hat{\boldsymbol{k}}) \hat{\boldsymbol{k}},0 \right)$		\\
	  & &	\\
	 	\hline
\;\;\;\;\;CMW &	$  \omega_{5}^{(0)}(\boldsymbol{k})=\,\left(\frac{\mathcal{C}}{\chi}-\frac{\mathcal{D}}{2}\frac{T^2}{\bar{w}}\right)\boldsymbol{B}.\boldsymbol{k}$ & $\delta \phi_{5}(k,\omega_{5}) = \left( 0,\boldsymbol{0},1 \right)$ \\
	\hline
	\caption{Hydrodynamic modes in a non-dissipative chiral fluid in presence of external magnetic field at $\bar{\mu}=0$. 
	}\label{tabelmodes3+1}
\end{longtable*}
In table (\ref{tabelmodes3+1diss}) we have re-expressed the modified dispersion relation of the  above modes when considering also the dissipative effects.
Just analogous to what was found in \cite{Abbasi:2015nka} for a neutral fluid, dissipation splits the degeneracy of CAWs here. Interestingly, the chiral  Alfv\'en is split into two chiral waves: one is a dissipating chiral  Alfv\'en wave and another is a dissipating mixed CMW/Alfv\'en wave.  
It should be also noted that the CMW changes to a mixed CMW/Alfv\'en due to dissipation.

\begin{longtable*}{|l|c|c|}
	\hline
Type of mode	&	Dispersion relation & $\sigma=\eta=\zeta=0$ \\
	\hline
   &	&	 \\
 \,\,\,\,\,\,\,\,\,\,\,\,\,sound   &	$\omega_{1,2}(\boldsymbol{k})=\,\pm v_{s} k-\,\frac{i}{2}\left(\boldsymbol{k}^2\gamma_{s}+\,\frac{\sigma}{ \bar{w}}\boldsymbol{B}^2 \sin^2 \theta \right)$&$\omega_{1,2}^{(0)}(\boldsymbol{k})$ \\
    &  &	\\
\hline
&   &	\\
\;\;\;\;\;\; Alfv\'en &	$\omega_{3}(\boldsymbol{k})=-\frac{\mathcal{D}}{2}\frac{T^2}{\bar{w}} \boldsymbol{B}.\boldsymbol{k}-i\left(\boldsymbol{k}^2 \gamma_{\eta}+\frac{\sigma}{ \bar{w}}\boldsymbol{B}^2 \cos^2 \theta
	\right)$  \,\,\,\,\,\,\,\,\,\,\,\,\, \,\,\,\,\,\,\,\,\,\,\,\,\, \,\,\,\,\,\,\,\,\,\,\,\,\, \,\,\,\,\,\,\,\,\,\,\,\,\, \,\,\,\,\,\,\,\,\,\,\,\,\, &	  $\omega_{3}^{(0)}(\boldsymbol{k})$\\
&   &	\\
mixed Alfv\'en/CMW   &   
	$\omega_{4,5}(\boldsymbol{k})=
	 \left(\frac{\mathcal{C}}{2\chi}-\frac{\mathcal{D}}{2}\frac{T^2}{\bar{w}}\right)\boldsymbol{B}.\boldsymbol{k}-\frac{i}{2}\left(  \boldsymbol{k}^2 (D+\gamma_{\eta})+ \frac{\sigma}{\bar{w}}\,\boldsymbol{B}^2 \right) $ \,\,\,\,\,\,\,\,\,\,\,\,\, \,\,\,\,\,\,\,\,\,\,\,\,\, \,\,\,\,\,\,\,\,\,\,\,\,\, \,\,\,\,\,\,\,\,\,\,\,\,\,& $\omega_{4}^{(0)}(\boldsymbol{k})$\\
& \,\,\,\,\,\,\,\,\,\,\,\,\, \,\,\,\,\,\,\,\,\,\,\,\,\, \,\,\,\,\,\,\,\,\,\,\,\,\, \,\,\,\,\,\,\,\,\,\,\,\,\, \,\,\,\,\,\,\,\,\,\,\,\,\,$\pm\frac{1}{2}\sqrt{\left(i \boldsymbol{k}^2(D-\gamma_{\eta})- i \frac{\sigma}{\bar{w}}\,\boldsymbol{B}^2- \frac{\mathcal{C}}{\chi}\boldsymbol{B}.\boldsymbol{k}\right)^2- \frac{4 D \sigma}{\bar{w}}\boldsymbol{B}^2\,\boldsymbol{k}^2\sin^2 \theta}$   	& $\omega_{5}^{(0)}(\boldsymbol{k})$\\
	\hline
	\caption{Hydrodynamic modes in presence of an external magnetic field at $\bar{\mu}=0$. In the expressions above, $\theta$ is the angle between momentum vector $\boldsymbol{k}$ and the magnetic field $\boldsymbol{B}$, i.e. $\cos\theta= \hat{\boldsymbol{k}}.\hat{\boldsymbol{B}}$.
	}\label{tabelmodes3+1diss}
\end{longtable*}
\subsection{Physical interpretation}
In \cite{Yamamoto:2015ria}, it has been discussed how an external magnetic field provides the restoring force needed for propagation of Chiral Alfv\'en waves in an anomalous charged fluid. We mention some more detail on how  the Lorentz force can make dissipation at the same time that it plays the role of a restoring force on the chiral current. We also discuss in this set up  only $\delta \phi_4$ propagates. 

Let us consider a chiral  Alfv\'en wave in an incompressible fluid.
As it can be seen in table \ref{tabelmodes3+1diss}, the magnetic field contributes through two terms in the expression of $\omega_3$: In the first term it leads to propagation of a chiral wave while in the last term it makes the propagating chiral wave dissipates. 

In order to understand how the magnetic field makes the role of a restoring force at the same time that it forces the wave to dissipate, we consider the following set up: let us take the magnetic field in the positive $z$-direction and consider the perturbation of fluid momentum\footnote{As was denoted earlier, the  Chiral Alfv\'en wave may propagate due to momentum fluctuations. } in the positive $y$-direction, $\boldsymbol{\pi}=\pi_y(z)\hat{\boldsymbol{y}}$ \cite{Yamamoto:2015ria}. As one expects, the momentum perturbation induces an Ohmic current as $\boldsymbol{J}_{\sigma}=\frac{\sigma}{\bar{w}} \boldsymbol{\pi} \boldsymbol{\times} \boldsymbol{B}$. In addition, since the fluid is chiral, the local vortical current $\boldsymbol{J}_{\omega}=\mathcal{D} T^2 \boldsymbol{\omega}$ is induced too. In presence of a  magnetic field, the above currents receive Lorentz forces $\boldsymbol{F}_{\sigma}$ and $\boldsymbol{F}_{\omega}$ respectively. 

\begin{figure}
	\begin{center}
			\includegraphics[scale=.65]{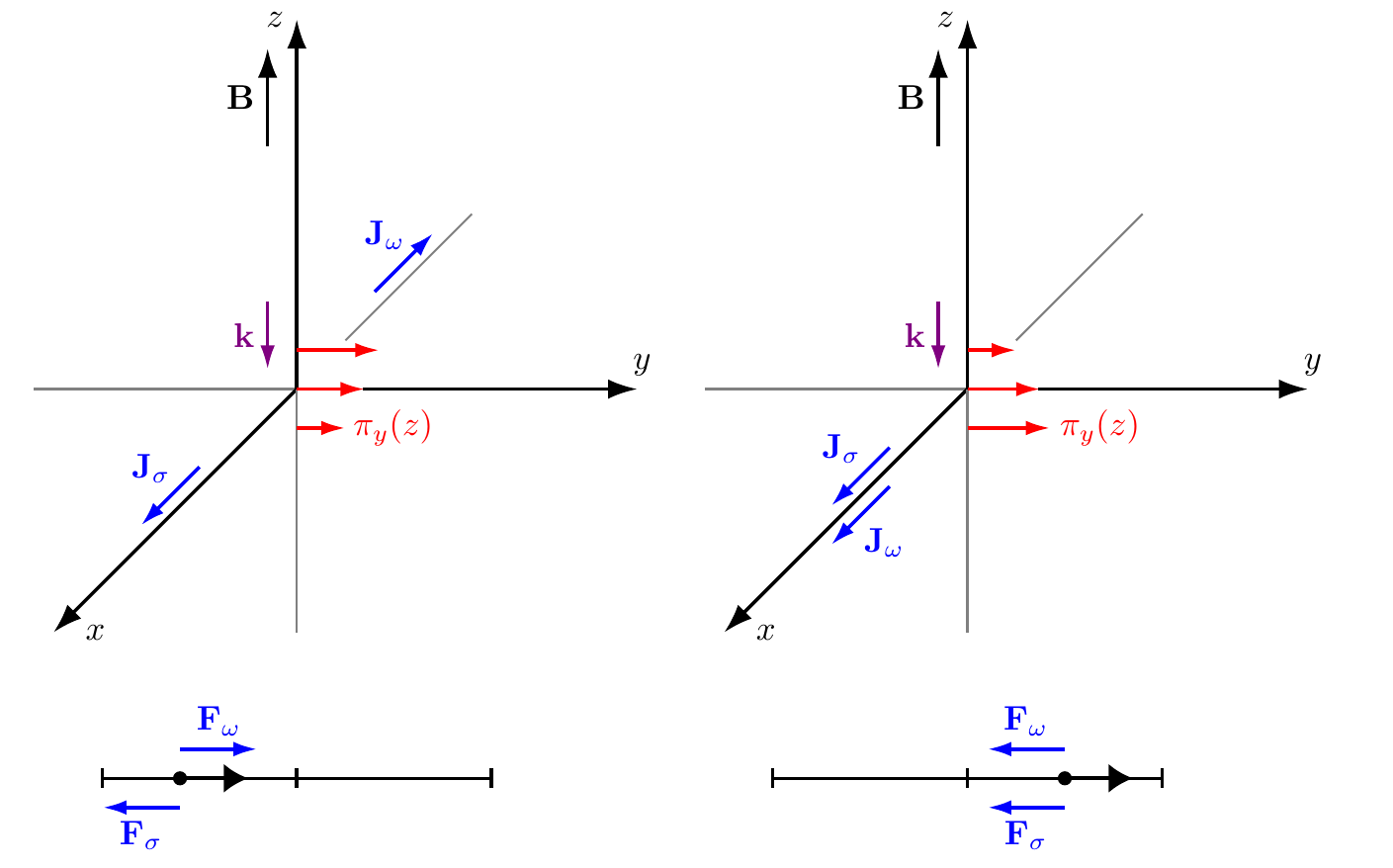}
	\end{center}
	\vspace{-0.7cm}
	\caption{Configuration of the vectors ${\bm B}$, ${\bm \pi}$, ${\bm \omega}$, ${\bm J}_{\sigma}$,
		${\bm J}_{\omega}$, ${\bm F}_{\sigma}$ and ${\bm F}_{\omega}$ for propagtion of $\delta\phi_4$}
	\label{fig:CAW}
\end{figure}

In Fig.\ref{fig:CAW} we have illustrated the Lorentz forces exerted on an element of the fluid at the origin in two different situations. 
In the left panel we assume  $\pi_y > 0$ and $\partial_z \pi_y > 0$. We also take the fluid incompressible, i.e. $\boldsymbol{\nabla}.\boldsymbol{v}=0$, so the wave-vector $\boldsymbol{k}$ points the negative $z$-direction (See $\omega^{\text{nd}}_3(\boldsymbol{k})$.).  For this local fluid momentum, the vorticity, $\boldsymbol{\omega}=\boldsymbol{\nabla}\times \boldsymbol{v}$, points along the negative $x$-direction. The resultant currents $\boldsymbol{J}_{\omega}$ and $\boldsymbol{J}_{\sigma}$ have been shown in the figure. At the bottom part of the figure, we have  illustrated the fluid element as a point oscillating on the $y$-axis near the origin, with the Lorentz forces exerted on it. At the situation considered above, the momentum of the element is increasing. This is due to both shape of the momentum profile and the direction of wave propagation; so at this moment, the fluid element behaves like an oscillator going towards the center of oscillation from the left hand side of it. As it can be clearly seen in the figure, $\boldsymbol{F}_{\omega}$ plays the role of the restoring force while  $\boldsymbol{F}_{\sigma}$ makes the momentum of the element dissipates.

Let us consider another situation in which $\boldsymbol{F}_{\omega}$ and $\boldsymbol{F}_{\sigma}$ act on the element in the same direction. To proceed, in the right panel of Fig.\ref{fig:CAW} we take $\pi_y > 0$ and $\partial_z \pi_y <0$. Compared to the left panel case, the direction of vorticity is reversed here. As a result, $\boldsymbol{J}_{\omega}$ and $\boldsymbol{F}_{\omega}$ are reversed too. Since the momentum of the element is decreasing now, it behaves like an oscillator going away from the center. Although both $\boldsymbol{F}_{\omega}$ and $\boldsymbol{F}_{\sigma}$ act in the same direction, the former is a restoring force pointing to the center and the latter is a friction-like force pointing opposite to the velocity.

In the set up considered above, we took the  wave vector along the direction of magnetic field. From the table (\ref{tabelmodes3+1}), it is obvious that the CAW with amplitude $\delta \phi_3$ will not be excited in this situation ($\hat{\boldsymbol{B}} \times \hat{\boldsymbol{k}}=0$). However the amplitude $\delta \phi_4$ propagates. The reason is as follows; consider $\theta$ as the angle between the wave vector and the direction of magnetic field. In order to study the $\hat{\boldsymbol{B}}\parallel \hat{\boldsymbol{k}}$ case in the expression of $\delta \phi_4$, we take the  wave vector fixed in the space and so rotate the magnetic field around it that $\theta \rightarrow 0$.    Having
\begin{equation}
\lim_{\theta \rightarrow 0} \big(\hat{\boldsymbol{B}}-(\hat{\boldsymbol{B}}.\hat{\boldsymbol{k}}) \hat{\boldsymbol{k}}\big)=\,\lim_{\theta \rightarrow 0}\big(\boldsymbol{\mathcal{R}(\theta)}\hat{\boldsymbol{k}}- \boldsymbol{1} \cos \theta \hat{\boldsymbol{k}}\big)=\,\boldsymbol{1}
\end{equation}
with $\mathcal{R}(\theta)$ the rotational matrix, simply confirms why when $\hat{\boldsymbol{B}}\parallel \hat{\boldsymbol{k}}$, the amplitude $\delta \phi_4$ propagates. 

An interesting point about $\delta \phi_3$ is that even if the magnetic field was transverse to the wave vector, this mode could not propagate. The reason for that  is in this case no restoring force exists to make $\delta \phi_3$ propagate. Mathematically, it is  obvious that in this limit ($\boldsymbol{B}.\boldsymbol{k}\rightarrow 0$),  $\omega_3$ vanishes .
It means that $\delta \phi_3$ may propagate in  every magnetic field except in directions either parallel or transverse to it.
  
In summary, while there could exist only two propagating wave in a normal fluid (sound modes), a neutral chiral fluid may have five hydrodynamic waves. To excite the three new waves, an external magnetic field is needed. The external magnetic field affects on an anomalous fluid through two ways: 1) providing the restoring force for propagation of chiral waves,  2) making dissipation via inducing Ohmic currents.

\section{Non-vanishing Chemical Potential}\label{sec4}
Now, let us consider a fluid of single right-handed fermions at finite density, namely at finite chiral chemical potential. 
Up to first order in derivative expansion, 
the linearized equations would be at most to second order in derivatives and take the following form:
\begin{eqnarray}\label{eqmotionchemic}
&\partial_t \delta \epsilon + \partial_i \pi_i = 0, \nonumber\\
 &\partial_t \pi_i +\beta_1 \partial_i \delta\epsilon  +\beta_2 \partial_i \delta n - \frac{\bar{n}}{\bar{w}} \, \epsilon^{ijl} \pi^j B^l - \frac{\xi}{2\bar{w}}\left( B^l\partial_l \pi^i - B^l \partial^i \pi_l \right) =    0,\\
 &\partial_t \delta n +\frac{\xi_B}{\bar{w}} B_i \partial_t \pi_i + \frac{\bar{n}}{\bar{w}} \; \partial_i \pi_i + B_i \;\left[ \left(\frac{\partial \xi_B}{\partial \epsilon}\right)_n  \partial_i \,\delta \epsilon+ \left(\frac{\partial \xi_B}{\partial n}\right)_\epsilon \partial_i \, \delta n \right] = 0, \nonumber
\end{eqnarray}
where $\bar{w}=\bar{\epsilon}+\bar{p}$ is the value of equilibrium enthalpy density. We have
also used:
\begin{equation}
\begin{aligned}
\delta p &= \beta_1 \delta\epsilon+ \beta_2 \delta n, \\
\delta \xi_B &= \left(\frac{\partial \xi_B}{\partial \epsilon}\right)_n \delta \epsilon+ \left(\frac{\partial \xi_B}{\partial n}\right)_\epsilon\delta n.
\end{aligned}
\end{equation}
In momentum space  equations \ref{eqmotionchemic} may be written as:
\begin{gather}
- i\omega \delta \epsilon + i k_i \pi_i = 0, \nonumber\\
+\beta_1 ik_i \delta\epsilon -i\omega \pi_i -  \frac{\bar{n}}{\bar{w}} \, \epsilon^{ijl} \pi_j B^l - \frac{\xi}{2\bar{w}}\left( B^l ik_l \pi_i - B^l ik_i \pi_l \right) +\beta_2 ik_i \delta n = 0,\\
 B_i \; \left(\frac{\partial \xi_B}{\partial \epsilon}\right)_n  ik_i \,\delta \epsilon + \frac{\bar{n}}{\bar{w}} \; ik_i \pi_i -\frac{\xi_B}{\bar{w}} B_i i\omega \pi_i - i\omega \delta n + B_i \left(\frac{\partial \xi_B}{\partial n}\right)_\epsilon ik_i \, \delta n = 0. \nonumber
\end{gather}
Due to presence of a term including $\omega\pi$, it would not be possible to restate the equations above in the matrix form analogous to \ref{eqsuperfield}. Instead, we may express them as:
\begin{equation}\label{eq:lin_mat}
 M_{ab}(\boldsymbol{k} , \omega)  \delta \phi_a (\boldsymbol{k} , \omega) = 0,
\end{equation}
with $M_{ab}$ given as:
\begin{widetext}
\begin{equation}
 M_{ab} = 
 \left( {\begin{array}{ccc}
- i\omega & i k_j & 0 \\
   \beta_1 ik^i & -i\omega \delta^i_j -i \frac{\xi}{2\bar{w}} \left(\boldsymbol{B} \cdot \boldsymbol{k} \delta^i_j - B_j k^i \right) - \frac{\bar{n}}{\bar{w}} \epsilon^i \,_{jl}B^l &  \beta_2 ik_i \\
   \left(\frac{\partial \xi_B}{\partial \epsilon}\right)_n  i \boldsymbol{B} \cdot \boldsymbol{k} & \frac{\bar{n}}{\bar{w}} i k_j - \frac{\xi_B}{\bar{w}} i\omega B_j & -i\omega +  \left(\frac{\partial \xi_B}{\partial n}\right)_\epsilon i \boldsymbol{B} \cdot \boldsymbol{k}
\end{array} } \right). 
\end{equation}
\end{widetext}
For \eqref{eq:lin_mat} to have a non-trivial answer, it is necessary to request
\begin{equation}\label{eq:det_M}
\det M_{ab}=0.
\end{equation}
This equation simply  gives the  dispersion relation of hydrodynamic excitations. In next subsections, we firstly study the hydrodynamic modes of a charge fluid at finite chemical potential in presence of a background magnetic field and in the absence of global anomalies. Since then we enter anomalies and study the effect of quantum anomalies on the hydrodynamic regime of a charged fluid at finite chiral chemical potential.

\subsection{The hydrodynamic modes in the absence of anomalies}
 The equations that yield the hydrodynamic modes and their associated amplitudes in this regime become:
\begin{equation}\label{eq:no_anomaly_det_M}
\begin{aligned}
\left( M_{ab}(\boldsymbol{k} , \omega)\big|_{\xi=0,\,\xi_B=0} \right) \delta \phi_a (\boldsymbol{k} , \omega) = 0,\\
 \det\left( M_{ab}\big|_{\xi=0,\,\xi_B=0}\right)=0.
\end{aligned}
\end{equation}
In what follows we take $B_{\perp}$ and $B_{\parallel}$ as the components of magnetic field orthogonal and parallel to the wave vector $\hat{k}$, respectively.
We divide the study of collective motions to three cases:
\begin{itemize}
\item $B_{\parallel} \neq 0 \, , \, B_{\perp} = 0$

Let us denote that the only objects we have in this case are two parallel vectors; so we are not able to present the amplitudes in a covariant way.  We freely take the wave-vector and the magnetic field both along the $z-$axis. By this choice,  the dispersion  relation of modes and their amplitudes may be given as it can be seen in table (\ref{tabelmodesChemicNoAnomaly}). (We have defined $\bar{\beta}= \beta _2 \bar{n}+\beta _1 \bar{w}$.)
\begin{longtable*}{|l|c|}
	\hline
\,\,\,\,\,\,\,\,\,\,\,\,\,\,\,Mode	& Eigen Vector   \\
	\hline
	\hline
\,\,\,\,\,\,\,\,\,\,\,\,\,\,\,\,$\omega_{1}^{(0)}=0$  & \,\,\,\,$\delta \phi_1(k,\omega_1) = \left(-\frac{\beta_2}{\beta_1},\,0,\,0,\,0,\,1 \right)$  \\
   &	\\
	\hline
	& \\
\,\,\,$\omega_{2,3}^{(0)} = \pm \frac{\bar{n} B_{\parallel}}{\bar{w}} = \pm \frac{\bar{n} B}{\bar{w}}$
	 \,\,\,&$
	 \delta \phi_{2,3}(k,\omega_{2,3}) = \left( 0,\,\pm i,\,1,\,0,\,0 \right),\nonumber$  	  \\	
	 & \\
	 \hline
	 & \\
\,\,\,\,\,\,\,\,$\omega_{4,5}^{(0)} = \pm k \sqrt{\frac{\bar{\beta}}{\bar{w}}}$ &	\,\,\,\,\,$  \delta \phi_{4,5}(k,\omega_{4,5}) = \left( 1,\,0,\,0,\,\pm\sqrt{\frac{\bar{\beta}}{\bar{w}}},\, \frac{\bar{n}}{\bar{w}}\right)$\\
& \\
	\hline
	\caption{Hydrodynamic modes in a non-dissipative chiral fluid in presence of external longitudinal  magnetic field $B_{\parallel}$ at $\bar{\mu}\ne0$. 
	}\label{tabelmodesChemicNoAnomaly}
\end{longtable*}
Among the non-zero modes given in the table, let us first interpret $\omega_{2,3}$. The frequency of these modes is obviously independent of the wave vector; so they are non-propagating modes. Since $\hat{\boldsymbol{k}}.\delta \phi_{2,3}=0$, they  just represent two   circularly polarized standing waves of the transverse momenta.  The presence of such vortex like rotating modes is the consequence of exerting the Lorentz force on the transverse momenta. This is a specific feature of the charged fluid and can not be observed in a neutral fluid even in presence of magnetic field. In next subsection we show that when $\bar{n}$ specifies the density of a chiral charge, these standing modes may both propagate due to effect of the anomalies. In fact $\omega_{2,3}$ found in the current case, are nothing but the gap between the chiral Alfv\'en waves in a charged fluid. 

Now let us consider modes  $\omega_{4,5}$. These are simply the longitudinal sound modes ($\hat{\boldsymbol{k}} \parallel \delta \phi_{4,5} $) whose velocity differs from the sound velocity in a neutral fluid ($v_s$):
\begin{equation}\label{speadsoundchemic}
\omega_{4,5}=\pm k \sqrt{\frac{\bar{\beta}}{\bar{w}}}=\pm k \sqrt{\beta_1+\frac{\beta_2 \bar{n}}{\bar{w}}}=\pm k \sqrt{v_s^2+\frac{\beta_2 \bar{n}}{\bar{w}}}
\end{equation}
 As it can be clearly seen, the speed of sound in this case, has nothing to do with the parallel magnetic field and  the difference relative to $v_s$ is only due to presence of non-vanishing charge in the fluid.  It could be also simply understood by considering the fluctuation of the pressure:
\begin{equation}
\delta p=\beta_1 \delta \epsilon+\beta_2 \delta n=\left(\frac{\partial p}{\partial \epsilon}\right)_n \delta \epsilon+\left(\frac{\partial p}{\partial n}\right)_{\epsilon} \delta n.
\end{equation}
In the limit where $\bar{n}=0$ the second term vanishes and only the energy perturbation contributes to propagation of sound. However, in the non-vanishing charge limit, the second term leads to appearance of a new contribution in the square root in (\ref{speadsoundchemic}). In another word  charge density perturbations may produce pressure gradient, as well as energy density perturbations do.

\begin{center}
\rule{10cm}{0.4pt}
\end{center}
\item $B_{\parallel} = 0 \, , \, B_{\perp} \neq 0$

To evaluate  the amplitude of fluctuations in this case, we first  take the wave-vector along the $z-$axis and the magnetic field  along the $y$ one. In this special frame we obtain: 
\begin{gather}\label{ampBtrans}
\delta \phi_a(k,\omega_1) = \left(-\frac{\beta _2}{\beta _1},0,0,0,1 \right),\\
\delta \phi_a(k,\omega_2) = \left( 0,0,1,0,0 \right),\nonumber\\
\delta \phi_a(k,\omega_3) = \left(-\frac{i \bar{n} }{\beta _1 \bar{w} },\frac{k}{B},0,0,0 \right), \nonumber\\
\delta \phi_a(k,\omega_{4,5}) = \left(\bar{w} ,- i \bar{n} \frac{B}{k},0,\pm \frac{1}{k} \sqrt{\bar{n}^2 B^2+\bar{w} k^2 \bar{\beta}},\bar{n} \right). \nonumber
\end{gather}
As we have shown in table  (\ref{tabelmodesChemicNoAnomalyBtrans}), the only non-zero modes in this case are $\omega_{4,5}$. With two elliptic polarization in the $x-z$ plane, these modes propagate in the $z-$direction.  As a result, these modes are neither transverse nor longitudinal. 
Similar to the sound wave, the propagation of $\pi_z$ perturbations along $z$ is due to the pressure gradient. However the magnetic field (directed along $y$ direction) exerts a  Lorentz force on the $\pi_{x}$ perturbation which makes the role of an extra pressure gradient in the $z$-direction. In order to mathematically show this increase in the sound speed,
let us rewrite the last two dispersion relations with the substitution $|\boldsymbol{B}| = \alpha_{\perp} k$:
\begin{gather}
\omega_{4,5} = \pm k \frac{1}{\bar{w}}\sqrt{\bar{n}^2 \alpha_{\perp}^2+\bar{w}\bar{\beta}  }. \nonumber
\end{gather}
Obviously, the velocity of these modes is greater than that of sound:
\begin{equation}\label{sound Chemic trans}
 v=\sqrt{v_s^2+\frac{\beta_{2}\bar{n}}{\bar{w}}+\frac{\bar{n}^2 \alpha_{\perp}^2}{\bar{w}^2}}.
\end{equation}
It can be seen that in addition to the pure contribution of the finite density, namely the term $\beta_2 \bar{n}/\bar{w}$ in the square root, there exists another contribution.  The origin of the term $\bar{n}^2\alpha_{\perp}^2/\bar{w}^2$ in the square root is the Lorentz force, discussed in previous paragraph, exerting on the fluid.

\begin{longtable*}{|l|c|}
	\hline
\,\,\,\,\,\,\,\,\,\,\,\,\,\,\,\,\,\,\,\,\,\,\,\,\,\,\,\,\,\,Mode	& Eigen Vector   \\
	\hline
	\hline
	&\\
  & \,\,\,\,$\delta \phi_1(k,\omega_1) = \left(-\frac{\beta _2}{\beta _1},\boldsymbol{0},1 \right)$,\\ \,\,\,\,\,\,\,\,\,\,\,\,\,\,\,\,\,\,\,\,\,\,\,\,\,\,\,\,\,\,\,$\omega_{1,2,3}^{(0)}=0$ &
$\delta \phi_2(k,\omega_2) = \left( 0,\frac{\boldsymbol{B}}{B},0 \right),\nonumber$\\ &
$\delta \phi_3(k,\omega_3) = \left(-\frac{i \bar{n}}{\beta _1 \bar{w} },\frac{\boldsymbol{B} \times \boldsymbol{k}}{B^2},0 \right)$\\
   &	\\
	 \hline
$\,\,\,\,\omega_{4,5}^{(0)} = \pm \frac{1}{\bar{w}}\sqrt{\bar{n}^2 B^2+\bar{w}\bar{\beta} k^2 } $\,\,\,\, &	\,\,\,\,\,$  \delta \phi_{4,5}(k,\omega_{4,5}) = \left(\bar{w} , - i \bar{n} \  \frac{\boldsymbol{B} \times \boldsymbol{k}}{k^2} \ \pm \ \frac{\boldsymbol{k}}{k^2} \sqrt{\bar{n}^2 B_{\perp}^2+\bar{w} k^2 \bar{\beta}},\bar{n}  \right)$\\
	\hline
	\caption{Hydrodynamic modes and their $SO(3)$ covariant amplitudes in a non-dissipative chiral fluid in presence of external transverse  magnetic field $B_{\perp}$ at $\bar{\mu}\ne0$. 
	}\label{tabelmodesChemicNoAnomalyBtrans}
\end{longtable*}
\begin{center}
\rule{10cm}{0.4pt}
\end{center}
\item $B_{\parallel} \neq 0 \, , \, B_{\perp} \neq 0$

Let us first consider the wave-vector is taken along $z$-axis  and the magnetic field has two components $B_\perp$ and $B_\parallel$ along axes $y$ and $z$, respectively. The corresponding amplitudes take the following form in this frame:
\begin{gather}
\delta \phi_a(k,\omega_1) = \left(-\frac{\beta _2}{\beta _1},0,0,0,1 \right),\\
\delta \phi_a(k,\omega_i) =  \left( \bar{w},\frac{-i \bar{n} \ B_{y} \, \omega_i^2}{k \left( \omega_i^2 - \bar{n}^2B_{z}^2 /\bar{w}^2\right)},\frac{-2\, \bar{n}^2 \, B_{y} B_{z} \, \omega_i/\bar{w}}{k \left( \omega_i^2 - \bar{n}^2B_{z}^2/\bar{w}^2 \right)},\frac{ \omega_i\bar{w}}{  k}, \bar{n} \right) , \qquad i =2,3,4,5. \nonumber
\end{gather}
Considering the following definitions, we have listed the dispesion relation  of the modes in addition to their covariant amplitudes in table (\ref{tabelmodesChemicNoanomalyboth B}).
\begin{gather}\label{eq:delta_a}
a = \bar{n}^2 B^2 + \bar{w} k^2 \bar{\beta},\\
\Delta =a^2-4 \bar{n}^2 \bar{w} (\boldsymbol{B}\cdot  \boldsymbol{k})^2 \bar{\beta} \nonumber
\end{gather}
Obviously, each of the non-zero modes is a mixture of longitudinal and transverse propagation.
However it is clear that generally, none of these mode are sound. All of them are modified sound modes which propagate with elliptic polarization in a plane neither parallel nor transverse to the wave vector.  Only in a special case where $B_{\perp}\rightarrow 0$, two of these modes become the ordinary longitudinal sound waves. 
It is also worth mentioning that  for any direction of the magnetic field, one of the five possible hydrodynamic modes will never be excited in the fluid. In next subsection we show that anomaly effects may excite the fifth  mode.
\end{itemize}

\begin{longtable*}{|l|c|}
	\hline
\,\,\,\,\,\,\,\,\,\,\,\,\,\,\,\,\,\,\,\,\,\,\,\,\,\,\,\,\,\,Mode\,\,\,\,\,\,\,\,\,\,\,\,\,\,\,\,\,\,	& Eigen Vector   \\
	\hline
	\hline
\,\,\,\,\,\,\,\,\,\,\,\,\,\,\,\,\,\,\,\,\,\,\,\,\,	$\omega_{1}^{(0)}=0$ & $\delta \phi_1(k,\omega_1) = \left(-\frac{\beta _2}{\beta _1},0,0,0,1 \right) $ \\
	\hline
	&\\
  \,\,\,\,\,\,\,\,\,\,$ \omega_{i}^{(0)} =
   \pm \frac{\sqrt{a+\pm\sqrt{\Delta}}}{\sqrt{2} \bar{w}} $ \,\,\,\,\,\,\,\,\,\,\,\,\,\,\,\,\,\,\,\, &	\,\,\,\,\,\,\,\,\,\,$\delta \phi_{2,3,4,5}(k,\omega_i) = \left( \bar{w}\, , \, \frac{-i \bar{n} \ \boldsymbol{k} \times \boldsymbol{B} \omega_i^2 - 2\frac{\bar{n}^2}{\bar{w}} \left( \boldsymbol{B} \cdot \boldsymbol{k} \right) \left( \boldsymbol{B} - (\boldsymbol{B} \cdot \boldsymbol{k}) \frac{\boldsymbol{k}}{k^2} \right) \omega_i }{\left( \omega_i^2 k^2 - \bar{n}^2 \left(\boldsymbol{B} \cdot \boldsymbol{k}\right)^2 /\bar{w}^2\right)} + \frac{ \omega_i\bar{w}}{ k^2} \boldsymbol{k} \, , \, \bar{n} \right) 
    , $\,\,\,\,\,\,\,\,\,\,\\
 \,\,\,\,\, $\qquad i =2,3,4,5$ &  $\qquad i =2,3,4,5$\\
 &\\     
	 \hline
	\caption{Hydrodynamic modes in a non-dissipative charged fluid in presence of external general magnetic field. 
	}\label{tabelmodesChemicNoanomalyboth B}
\end{longtable*}
\begin{center}
\rule{10cm}{2pt}
\end{center}

\subsection{Anomalous fluid with finite chiral chemical potential}
In this subsection we are going to take the effect of anomalies into account.  It is important to note that the effect of anomalies enters via the non-anomalous transport coefficients. These coefficients appear at the first order of hydrodynamic derivative expansion.
So all what we compute in this subsection will be to compute the derivative type corrections to our results in previous subsection.  We limit our study to computing the derivative corrections to the dispersion relations. Just analogue of what we did in case of charged fluid with no anomaly, we divide our current study into three different cases.
\begin{center}
\rule{10cm}{2pt}
\end{center}
\begin{itemize}
\item $B_{\parallel} \neq 0 \, , \, B_{\perp} = 0$

Using the modes given in table (\ref{tabelmodesChemicNoAnomaly}) as the zero order solution to the equation (\ref{eq:det_M}), we find the collective excitations to first order as the following:
\begin{equation}\label{anomalymodeparallel}
\begin{split}
\omega_1 =& \frac{\bar{w}}{\bar{\beta}}\left(\beta _1 \left(\frac{\partial \xi_B}{\partial n}\right)_\epsilon - \beta _2 \left(\frac{\partial \xi_B}{\partial \epsilon}\right)_n\right) B k \\
\omega_{2,3} =&\pm \frac{\bar{n} }{\bar{w}}B - \frac{\xi  }{2 \bar{w}}B k\\
\omega_{4,5} = &\pm \frac{1}{\bar{w}}\sqrt{\bar{n}^2 B^2+\bar{w}\bar{\beta} k^2 } +\frac{\beta _2}{2 \bar{\beta}}\left(\bar{w} \left(\frac{\partial \xi_B}{\partial \epsilon}\right)_n + \bar{n} \left(\frac{\partial \xi_B}{\partial n}\right)_\epsilon - \frac{\xi _B}{\bar{w}} \bar{\beta}\right) B k 
\end{split}
\end{equation}
Let us remind that $B, k \sim O(\partial)$ and to first order in derivative expansion of hydrodynamic constitutive relations, dispersion relations would normally include terms with at most two derivatives.

The first mode $\omega_1$ denotes the CMW in a chiral charged fluid. Compared to CMW in neutral chiral fluid given in (\ref{first CMW}), we find a new contribution as the following:
\begin{equation}
-\frac{\bar{w}}{\beta}\,\beta_2 \left(\frac{\partial \xi_B}{\partial \epsilon}\right)_n\,B k=\,-\frac{\beta_2 \bar{n}}{2 \beta \bar{w}}\left(\mathcal{C}\mu^2+\frac{}{}\mathcal{D}T^2\right) B k.
\end{equation} 
The next two modes, namely $\omega_{2,3}$, are nothing but CAWs. As we explained in previous subsection, the net chiral charge of the fluid makes the CAW gapped. The most interesting feature of the results might be related to the last two sound modes. We remember from the case of a neutral chiral fluid 
that the sound modes are not affected by the effect of anomaly. However when the fluid is chirally charged, from the last line of \ref{anomalymodeparallel} it is observed that the sound modes become dispersive. It would be better seen when writing the modes with substituting  $|\boldsymbol{B}| = \alpha_{\parallel} k$:
\begin{gather}
\omega_1 =\frac{\bar{w}}{\bar{\beta}}   \left(\beta _1 \left(\frac{\partial \xi_B}{\partial n}\right)_\epsilon - \beta _2 \left(\frac{\partial \xi_B}{\partial \epsilon}\right)_n\right) \alpha_{\parallel} k^2 ,\\
\omega_{2,3} = \pm  \frac{\bar{n}}{\bar{w}}  \alpha_\parallel k - \frac{\xi  }{2 \bar{w}} \alpha_\parallel k^2,\nonumber\\
\omega_{4,5} = \pm  \sqrt{\frac{\bar{\beta}}{\bar{w}}} k +  \frac{\beta _2}{2 \bar{\beta}} \left(\bar{w} \left(\frac{\partial \xi_B}{\partial \epsilon}\right)_n + \bar{n} \left(\frac{\partial \xi_B}{\partial n}\right)_\epsilon - \frac{\xi _B}{\bar{w}} \bar{\beta}\right)  \alpha_\parallel k^2. \nonumber
\end{gather}
The dispersive part of the sound only exists when the chiral density is finite. 
\begin{center}
\rule{10cm}{0.4pt}
\end{center}
\item $B_{\parallel} = 0 \, , \, B_{\perp} \neq 0$

The values of the frequencies do not differ from those of a non-anomalous fluid in this case (See table (\ref{tabelmodesChemicNoAnomalyBtrans})). It means that anomaly effects can not be detected in directions transverse to the magnetic field even if the fluid is chirally charged. In another word, no first order parity odd correction contributes to the collective excitations in the direction transverse to the magnetic field.
\begin{center}
\rule{10cm}{0.4pt}
\end{center}
\item $B_{\parallel} \neq 0 \, , \, B_{\perp} \neq 0$

In this part we  give the results corresponded to propagation of hydrodynamic waves in  an arbitrary direction with respect to an external magnetic field. At zero order in derivative expansion, there exist two modified sound excitations ($\omega_{4,5}$) in addition to another two mixed transverse-longitudinal waves($\omega_{2,3}$): 
\begin{gather}
\omega^{(0)}_1 = 0,\\
\omega^{(0)}_{2,3} = \pm \frac{\sqrt{-\sqrt{\Delta}+a}}{\sqrt{2} \bar{w}}, \nonumber\\ 
\omega^{(0)}_{4,5} = \pm \frac{\sqrt{\sqrt{\Delta}+a}}{\sqrt{2} \bar{w}}. \nonumber 
\end{gather}
Where $\Delta$ and $a$ are defined in \eqref{eq:delta_a}.
The fully covariant corrections in first order are given in table (\ref{tabelmodesGeneral}).
\begin{longtable*}{|l|c|}
	\hline
\,\,\,\,\,\,\,\,\,\,Zero Order	& First Order   \\
	\hline
	\hline
	& \\
	\,\,\,\,\,\,\,\,\,\,$\omega^{(0)}_1 = 0$ \,\,\,\,\,\,\,\,&  $\omega^{(1)}_1 = \frac{\bar{w}}{\beta} \left(\boldsymbol{B} \cdot \boldsymbol{k}\right) \left(\beta _1 \left(\frac{\partial \xi_B}{\partial n}\right)_\epsilon - \beta _2 \left(\frac{\partial \xi_B}{\partial \epsilon}\right)_n\right)$\\
	& \\
	\hline
	&
	\\
\,\,\,$\omega^{(0)}_{2,3} =  \pm \frac{\sqrt{a-\sqrt{\Delta}}}{\sqrt{2} \bar{w}}$\,\,\,\,& \,\,\,\,$\omega^{(1)}_{2,3}= 
\frac{\left(\boldsymbol{B} \cdot \boldsymbol{k}\right)}{4 \sqrt{\Delta } \bar{w}^2 } \left( - \left(2 \bar{w}^2 k^2 -4 \bar{n}^2 \bar{w}^2 \frac{\left( \boldsymbol{B} \cdot \boldsymbol{k} \right)^2}{a-\sqrt{\Delta}}\right) \left(\beta _2 \left(-\bar{\beta} \xi _B+\bar{w}^2 \left(\frac{\partial \xi_B}{\partial \epsilon}\right)_n + \bar{n} \bar{w} \left(\frac{\partial \xi_B}{\partial n}\right)_\epsilon\right)+\bar{\beta} \xi \right)-2 \sqrt{\Delta } \xi  \bar{w} \right)$\,\,\,\,\\  
   & \\
	 \hline
	 &
	 \\
\,\,\,\,$\omega^{(0)}_{4,5} =\pm \frac{\sqrt{a+\sqrt{\Delta}}}{\sqrt{2} \bar{w}}$\,\,\,\,& \,\,\,\,$\omega^{(1)}_{4,5} = \frac{  \left(\boldsymbol{B} \cdot \boldsymbol{k}\right)}{4 \sqrt{\Delta } \bar{w}^2}
 \left( \left(2 \bar{w}^2 k^2 -4 \bar{n}^2 \bar{w}^2 \frac{\left( \boldsymbol{B} \cdot \boldsymbol{k} \right)^2}{a+\sqrt{\Delta}}\right) \left(\beta _2 \left(-\bar{\beta} \xi _B+\bar{w}^2 \left(\frac{\partial \xi_B}{\partial \epsilon}\right)_n + \bar{n} \bar{w} \left(\frac{\partial \xi_B}{\partial n}\right)_\epsilon\right)+\bar{\beta} \xi \right)-2 \sqrt{\Delta } \xi  \bar{w} \right)$\,\,\,\,\\  
&\\
	\hline
	\caption{Hydrodynamic modes in a non-dissipative chiral fluid in presence of general external magnetic field at finite chemical potential. 
	}\label{tabelmodesGeneral}
\end{longtable*}
The presence  of a factor $\boldsymbol{B} \cdot \boldsymbol{k}$ in front of all $\omega^{(1)}$s is in agreement with our argument in  part ($B_{\parallel} = 0 \, , \, B_{\perp} \neq 0$) of the current subsection. Apart from the CMW $\omega_1$, there exist four mixed dispersive modes. These four modes are in general mixed Modified Sound-Alf\'ven waves.  In the special case $\boldsymbol{B} \parallel \boldsymbol{k}$ two of these modes become sound waves; the other two are CAWs appearing just from the first order. That two of these modes vanish at zero order can be simply understood  by considering this point that when $\boldsymbol{B} \parallel \boldsymbol{k}$, depending on whether $\bar{n} B$ is greater or $\bar{w} \bar{\beta} k^2$, one of the expressions $a+\sqrt{\Delta}$ and $a-\sqrt{\Delta}$ vanishes(See eqs (\ref{eq:delta_a})) while all  $\omega^{(1)}$s remain non-vanishing .
%
%
%
\end{itemize}

\begin{center}
\rule{10cm}{2pt}
\end{center}

\section{Parity violating fluid in $1+1$ dimensions}\label{sec5}
What all we have done so far was especially related to chiral fluids in 3+1 dimensions.
As it is well known, the chiral anomaly is also present in even space-time dimensions. Knowing this fact, chiral fluids have been also studied in 1+1 dimensions in the literature.  There are well known results concerning the anomalous
transport  in 1+1 dimensions found from both effective field theory \cite{Dubovsky:2011sk} and partition function \cite{Jain:2012rh} methods.

 Specifically, the authors of \cite{Dubovsky:2011sk} has considered a Wess-Zumino-like term to account the effect of anomalies. Interestingly, they have shown that in the spectrum of collective excitations of chiral fluid in 1+1 dimensions, in addition to  two ordinary sound modes, there exists
a new propagating mode; a right- or left-moving wave with propagation speed that goes to zero with the anomaly coefficient.  

Analogous to $3+1$ dimensions, the hydrodynamic spectrum can be found directly from the linearized hydrodynamic equations in $1+1$ dimensions. In what follows we study the linearized equations of chiral hydrodynamic in $1+1$ dimensions and show that there exist exactly three hydrodynamic modes as found in \cite{Dubovsky:2011sk}. Furthermore we rewrite the dispersion relation of each hydrodynamic mode in an explicit expression of  thermodynamic variables and the anomaly coefficient in the Landau-Lifshitz frame. Let us note that in \cite{Dubovsky:2011sk}, the dispersion relation of hydrodynamic modes have been given in the limit $c\rightarrow 0$ and  in the entropy frame  ($s^{\mu}=\bar{s} u^{\mu}$). 

The hydrodynamic equations for a chiral fluid in $1+1$ dimensional flat space-time in presence of an external long-wave length gauge filed read 
\begin{equation}
\begin{split}
\partial_{\mu}T^{\mu \nu}=&\,F^{\nu \lambda} J_{\lambda}\\
\partial_{\mu} J^{\mu}=&\,c\, \epsilon_{\mu \nu}F^{\mu \nu}
\end{split}
\end{equation}
with the anomaly coefficient $c$. The constitutive relations at zero order in derivative expansion are
 \begin{equation}\label{T J 1+1}
 	\begin{split}
 		T^{\mu \nu}=& \,(\epsilon+p) u^{\mu} u^{\nu}+ p \,\eta^{\mu \nu} \\
 		J^{\mu}=& \,n u^{\mu} +\xi \tilde{u}^{\mu}.
 	\end{split}
 \end{equation}
where $\tilde{u}^{\mu}=\epsilon^{\mu \nu}u_{\nu}$ and the coefficient $\xi$ appearing in front of the parity violating term is an anomalous transport coefficient, as \cite{Dubovsky:2011sk,Jain:2012rh} 
\begin{equation}
\xi=\,c\left(\frac{\bar{n} \mu^2}{\bar{\epsilon}+\bar{p}}- 2 \mu\right)-d\,\frac{\bar{n}T^2}{\bar{\epsilon}+\bar{p}}.
\end{equation}
 Let us recall that in $3+1$ dimensions, the anomaly effects arise from the first order in derivative expansion; that in $1+1$ dimensions there existed one anomalous coefficient even at zero order, is simply due to the rank of Levi-Civita tensor in $1+1$ dimensions. 
 
To find the spectrum of the fluid, we have to specify the state of equilibrium.
Since no magnetic field exists in $1+1$ dimensions, we take the state of equilibrium as 
\begin{equation}
u^{\mu}=(1,0),\,\,\,T=\text{Const.},\,\,\,\mu=0.
\end{equation}
 So the linearized hydrodynamic equations around the above state are
 \begin{equation}\label{linearized1+1}
 	\begin{split}
 		& \partial_{t} \delta \epsilon +\, i k \pi=0\\
 		& \partial_{t}\pi+\,i k v_s^2 \delta \epsilon=0\\
 		& \partial_{t}  n- \frac{\xi}{\bar{w}}\partial_t \pi+ i \left(\frac{\partial \xi}{\partial n}\right)_{\epsilon}  k\, n\,=0.\\
 	\end{split}
 \end{equation}
Analogue of what we did in the case of a chiral fluid in $3+1$ dimensions, we take the super filed $\pi_a=\big(\delta \epsilon, \pi, n \big)$ and rewrite the above linearized equations in the form  
$ \partial_{t} \phi_{a}(t, \boldsymbol{k})+\,M_{a b}(\boldsymbol{k})\, \phi_{b}(t, \boldsymbol{k})=0$ with
\begin{equation}
-i \omega \delta_{a b}+ M_{a b}(\boldsymbol{k})=
\left( {\begin{array}{ccc}
	- i \omega  & i k & 0\\ 
	i k^{i} v_{s}^2 & - i \omega & 0  \\
	0    & i \frac{\xi}{\bar{w}} \omega  & - i \omega + i \left(\frac{\partial \xi}{\partial n}\right)_{\epsilon} k 
	\end{array} } \right). 
\end{equation}
Equating the det($-i \omega \delta_{a b}+ M_{a b}(\boldsymbol{k})$) to zero, we find three hydrodynamic modes in an ideal chiral fluid in $1+1$ dimensions
\begin{equation}
\begin{split}
\omega_{1,2}(\boldsymbol{k})&=\, v_s k=\, \left(\frac{\partial p}{\partial \epsilon}\right) k\\
\omega_{3}(\boldsymbol{k})&=\,\left(\frac{\partial \xi}{\partial n}\right)_{\epsilon} k=\,\left(-\frac{2 c }{\chi}-\frac{d T^2}{\bar{w}}\right) k.
\end{split}
\end{equation} 
Obviously, the modes $\omega_{1,2}$ are the  sound modes. The third mode, namely $\omega_3$, must be the one-and-halfth sound mode earlier found through the effective filed theory approach.   

In summary,  a chiral fluid in $1+1$ dimensions has three propagating mode; two ordinary sound waves and one chiral wave. Compared to \cite{Dubovsky:2011sk}, we have computed the velocity of chiral wave in terms of anomaly coefficient and the value of thermodynamic variables in equilibrium. An undetermined integral constant, namely $d$, is also present in the dispersion-relation of this mode. 
٪٪٪٪٪٪٪٪٪٪٪٪٪٪٪٪٪٪٪٪٪٪٪٪٪٪٪٪٪٪٪٪٪٪٪٪٪٪٪٪٪٪٪٪٪٪٪٪٪

\section{Summary and Outlook}
In this paper we computed the spectrum of hydrodynamic fluctuations for a chiral fluid in the presence of an external magnetic field. As one naturally expect, five distinguished hydrodynamics modes may propagate in fluid with one $U(1)$ global symmetry in 3+1 dimensions. When the current is conserved, only two of the five modes may be excited by perturbing the fluid. These are nothing but the sound modes. We have shown that when the $U(1)$ current is anomalous, an external magnetic field would be able to turn on all five possible hydrodynamic excitations. In the limit of vanishing net chiral charge, the three new modes are as follow: a degenerate Chiral Alf\'en Wave and a  Chiral Magnetic Wave. We have also shown that the degeneracy between the CAWs might be removed if the effects of dissipation were considered.

When the fluid is chirally charged, similar to the previous case, five hydrodynamic modes may propagate. However the feature of propagation is somewhat different from that of happens in an uncharged chiral fluid. Here, sound waves combine with Chiral Alf\'en waves into for mixed waves. The new mixed waves are neither transverse nor longitudinal. The only unchanged mode is Chiral Magnetic Wave by this mean that it is a wave of scalar (density) perturbations yet, although its speed of propagation compared to the uncharged chiral fluid case. 

While the main outcome of our computations is that the anomaly effects may macroscopically appear through hydrodynamic waves in the magnetic field, it is worth mentioning that these waves may propagate in every arbitrary directions except in direction perpendicular  to the magnetic field.

What we have done in this work may be simply generalized to the case of a chiral fluid with both axial and vector currents. Whether in presence of two currents Chiral Alf\'ven wave propagates or not  is an interesting question which might be important in quark-gluon plasma physics. We leave more investigation on this issue to our future work.

In another direction, very recently  the author of  \cite{Chernodub:2015gxa}  has shown that  in a charged fluid at zero chemical potential a new type of chiral waves, different from the chiral Alfv\'en waves, may propagate. 
Analogous to the chiral Alfv\'en waves, the new wave, namely the chiral heat wave, is associated with  anomalous effects with this difference that for the latter, the necessary condition of propagation is presence of a background vorticity $\boldsymbol{\omega}$ in  the fluid. It would be interesting to investigate how the dissipative processes affect the chiral heat waves. It would be also interesting to study the mixing of chiral Alfv\'en/Heat waves in a fluid when considering both an external magnetic field and a constant vorticity in the fluid. 

Another interesting problem is to study the Chiral Alf\'ven waves in the holography. As it is well known, to every long wavelength perturbation of Einstein equations in Ads5 space, one fluid dynamical flow on the boundary of AdS is corresponded. The latter statement is the main subject of fluid/gravity duality   \cite{Bhattacharyya:2008jc}. So it might be possible to determine CAW  on the boundary of AdS is corresponded to which gravity set up in the AdS side. 

\section{Acknowledgements}
 We would like to thank to Naoki Yamamoto for exchanging two emails. We would also like to thank to Massimo Giovannini for discussion. K.H. wishes to thank institute for research in fundamental sciences, school of Particles and Accelerators for hospitality and partial financial support, and also acknowledges Hessamadin Arfaei for introducing him to the research group of A.D. and N.A.

\section*{Appendix- Comment on Parity violating fluid in $2+1$ dimensions}
\label{App}
As we discussed in the text, chiral anomaly is present only in even spaec-time dimensions. 
So the presence of parity violating terms in hydrodynamic currents in odd space-time dimensions can not be related to anomaly. However, it is instructive to investigate how these terms affect the hydrodynamic transport in odd spaec-time dimensions.

In \cite{Jensen:2011xb}, all transport coefficients of first order hydrodynamics in a parity broken system have been classified in $2+1$ dimensions. Furthermore, in the same paper the second law of thermodynamics, time reversal symmetry and properties of response functions have all been used to constrain the transport coefficients.

Due to absence of anomalies in $2+1$ dimensions, we do not expect the parity violating terms introduced in \cite{Jensen:2011xb} could make new propagating modes in the fluid.
However it would be instructive to study the effect of these terms on the hydrodynamic modes. To proceed, we repeat the computations of previous sections for a non-dissipative parity violating fluid in $2+1$ dimensions.

In $2+1$ dimensions, the first order corrections in (\ref{TJ})  are given by:
\begin{equation}
	\label{E:T1J1L}
	\begin{split}
	&   \tau^{\mu\nu} = \left( - \zeta \nabla_\alpha u^\alpha - \tilde{\chi}_B B - \tilde{\chi}_\omega \omega\right) \Delta^{\mu\nu}
	-\eta \sigma^{\mu\nu} - \tilde{\eta} \tilde{\sigma}^{\mu\nu} \,,  \\
	&  \nu^\mu = \sigma V^{\mu} + \tilde{\sigma} \tilde{V}^{\mu} + \tilde{\chi}_E \tilde{E}^{\mu} + \tilde{\chi}_Tbe \epsilon^{\mu\nu\rho}u_{\nu} \nabla_{\rho} T \,.
	\end{split}
\end{equation}
with
\begin{subequations}
	\label{E:defs}
	\begin{align}
	\label{E:OandB}
	& \omega = -\epsilon^{\mu\nu\rho}u_{\mu} \nabla_{\nu} u_{\rho}, \,\,\,\,\,\,\,\,\,
	 B = -\frac{1}{2} \epsilon^{\mu\nu\rho}u_{\mu} F_{\nu\rho}, \\
	& E^{\mu}  =  F^{\mu\nu}u_{\nu},\,\,\,\,\,\,\,\,\,
	V^{\mu}  = E^{\mu} - T p^{\mu\nu}\nabla_{\nu} \frac{\mu}{T}, \\
	\label{E:sigmaDef}
	& P^{\mu\nu} = u^{\mu}u^{\nu} + g^{\mu\nu}, \,\,\,\,\,\,\,\,\,\\
	&
	 \sigma^{\mu\nu}  = P^{\mu\alpha} P^{\nu\beta} \left(\nabla_{\alpha}u_{\beta} + \nabla_{\beta} u_{\alpha} - g_{\alpha\beta} \nabla_{\lambda} u^{\lambda} \right) \,,
	\intertext{and}
	&\tilde{E}^{\mu}  = \epsilon^{\mu\nu\rho}u_{\nu}E_{\rho}\,,\,\,\,\,\,\,\,\,\,
	\tilde{V}^{\mu}  = \epsilon^{\mu\nu\rho}u_{\nu} V_{\rho}\,, \\
	&\tilde{\sigma}^{\mu\nu} = \frac{1}{2} \left( \epsilon^{\mu\alpha\rho} u_{\alpha} \sigma_{\rho}^{\phantom{\rho}\nu} +  \epsilon^{\nu\alpha\rho} u_{\alpha} \sigma_{\rho}^{\phantom{\rho}\mu} \right)\,.&
	\end{align}
\end{subequations}
As before, the thermodynamic parameters $\bar{p}(\mu,T)$, $\bar{\epsilon}(\mu,T)$ and $\bar{n}(\mu,T)$ are the values of the pressure, energy density and charge density respectively in an equilibrium configuration in which $B=\omega=0$, where $B$ is the rest-frame magnetic field and $\omega$ the vorticity \footnote{In $2+1$ dimensions, both magnetic field and the vorticity are scalar quantities.}. 

Linearizing the hydrodynamic equation around the state
\begin{equation}
u^{\mu}=(1,0,0),\,\,\,T=\text{Const.},\,\,\,\mu=0,
\end{equation}
we find four hydrodynamic modes of which, two modes are zero modes which will become shear and heat modes when accounting the dissipative effects. The other two modes are the sound modes
\begin{equation}
\omega_{1,2 }(\boldsymbol{k})=\,\pm v_{s} k\left(1-\frac{T  B}{2 \bar{w}}\left(\tilde{\chi}_T+ c_v\frac{\partial \tilde{\chi}_B}{\partial \epsilon}\right)\right)
\end{equation}

%
The parameters $\tilde{\chi}_B$ and $\tilde{\chi}_T$ are not independent; indeed at $\mu=0$ they are specified in terms of a thermodynamic function
\begin{equation}
\label{E:MMO}
\mathcal{M}_{B} =  \frac{\partial P}{\partial B}
\end{equation}
and its derivatives with respect to $T$ and $\mu$ \cite{Jensen:2011xb}.
Our above result shows that the parity violating terms in $2+1$ dimensions can only affect  the sound propagation's speed in a magnetized fluid \cite{Abbasi:2015nka}. As a result no new propagating mode would appear due to parity breaking in $2+1$ dimensions. This result illustrates the importance of relation between anomalies and parity violating terms in even space-time dimensions discussed earlier.

\bibliographystyle{utphys}


\providecommand{\href}[2]{#2}\begingroup\raggedright\endgroup

\end{document}